\documentclass[journal=jpccck,manuscript=article]{achemso}

\usepackage{chemformula} 
\usepackage[T1]{fontenc} 

\usepackage{amsmath,txfonts,bm}
\usepackage{siunitx}
\usepackage{color}
\usepackage{ulem}
\usepackage{comment}

\author{Kazutaka Nishiguchi}
\affiliation[Kobe University]{Graduate School of System Informatics, Kobe University, 1-1 Rokkodai-cho, Nada-ku, Kobe 657-8501, Japan}
\email{nishiguchi@pearl.kobe-u.ac.jp}
\author{Ryota Yamamoto}
\affiliation[Kobe University]{Graduate School of System Informatics, Kobe University, 1-1 Rokkodai-cho, Nada-ku, Kobe 657-8501, Japan}
\author{Seiichiro Ten-no}
\affiliation[Kobe University]
{Graduate School of System Informatics, Kobe University, 1-1 Rokkodai-cho, Nada-ku, Kobe 657-8501, Japan}
\email{tenno@garnet.kobe-u.ac.jp}
\title{Theoretical Investigation of Oxygen-Vacancy Diffusion in SrTiO$_3$ using Neural Network Potentials}

\begin{document}

%
%
%
%
%

\begin{abstract}

Oxygen-vacancy formation and diffusion play important roles in determining the electronic and atomic transport properties of SrTiO$_3$ (STO). 
Accurate analysis of such point defects requires sufficiently large-scale and long-time molecular dynamics (MD) simulations; however, direct investigations of oxygen-vacancy dynamics using {\it ab initio} MD (AIMD) based on density functional theory (DFT) are limited to relatively small systems and short time scales because of their high computational cost. 
In this study, we develop a neural network potential (NNP) for pristine and oxygen-deficient STO using a reference dataset generated from AIMD simulations. 
The trained NNP model accurately reproduces the DFT reference energies, atomic forces, and virials for both pristine STO and STO containing an oxygen vacancy.
The NNP also reproduces the supercell-size dependence of the total energies and oxygen-vacancy formation energies, including the formation energy extrapolated to the dilute limit. 
Large-scale and long-time MD simulations using the NNP (NNP-MD) are then performed to investigate oxygen-vacancy diffusion. 
The resulting oxygen-vacancy diffusion coefficients show substantially better agreement with experimental data than those obtained from conventional classical MD simulations. 
An Arrhenius analysis yields an oxygen-vacancy migration enthalpy of 0.64 eV, in close agreement with the experimental value of 0.67 eV. 
These results demonstrate that NNPs trained on first-principles data provide an accurate and computationally efficient framework for investigating defect formation and long-time vacancy transport in perovskite oxides.

\end{abstract}


\section{Introduction}

Oxygen vacancies (V$_{\rm O}$) in perovskite oxide SrTiO$_3$ (STO) have been studied as a typical example of anion defects in perovskite compounds. 
Although pristine STO is considered an intrinsic semiconductor,~\cite{Piskunov_2004,Begum_2019} the synthesized samples often exhibit n-type semiconducting behaviors due to the presence of V$_{\text{O}}$. 
These oxygen defects make a significant effect on the electronic structure due to the formation of defect levels within the energy band gap since V$_{\text{O}}$ acts as an n-type doping and introduces excess electron carriers.~\cite{Hou_2010,Samanta_2012}  
The static electronic properties, including the density of states, band structure, and the formation energy of V$_{\text{O}}$, have been investigated using first-principles calculations based on density functional theory (DFT).~\cite{Nishiguchi_2025,Banerjee_2026} 
However, the dynamics and thermodynamics, such as oxygen vacancy diffusion and heat capacity, are difficult to address using molecular dynamics (MD) simulations: 
ab initio MD (AIMD) simulations are restricted to small-scale systems and short-time simulations due to their high computational costs, 
while the accuracy of classical MD simulations is limited despite their low computational costs.~\cite{Katsumata_1998,Schie_2012,Waldow_2016,Dawson_2020} 

In recent decades, machine learning interatomic potentials (MLIP), including neural network potential (NNP),~\cite{NNP_2007} Gaussian approximation potential (GAP),~\cite{GAP_2010} and spectral neighbor analysis potential (SNAP),~\cite{SNAP_2015} have attracted much interest because of their lower computational costs by several order of magnitude while achieving comparable accuracy with DFT calculations. 
These MLIPs have been successfully applied not only to molecules and solids but also to complex systems such as alloy,~\cite{Shimizu_2021} amorphous materials,~\cite{Andrade_2020} and interfaces.~\cite{Chaney_2024} 
MD simulations with the NNP model (NNP-MD) have also been applied to pristine STO to describe the structural phase transition between the cubic and tetragonal crystal phase.~\cite{He_2022} 
Currently, active learning (AL) techniques are being actively developed to reduce the amount of labeled data.~\cite{Smith_2018,Vandermause_2020,Sivaraman_2020,Zhang_2020,Shimizu_2021,Novikov_2021,Podryabinkin_2023,van_der_Oord_2023,Kulichenko_2023} 

This paper presents NNP-MD simulations of pristine and oxygen-deficient STO to investigate oxygen-vacancy formation energetics and oxygen-vacancy diffusion dynamics. 
The NNP model is trained using reference datasets generated from AIMD simulations and is validated through comparisons with DFT energies and atomic forces. 
The total energies of pristine STO and the oxygen-vacancy formation energies for various supercell sizes are predicted by the NNP model. 
The NNP predictions reproduce these static energetics including their extrapolated values in the dilute-defect limit. 
Large-scale and long-time NNP-MD simulations are performed to evaluate the mean-square displacements (MSDs) of the constituent atomic species and the oxygen-vacancy diffusion coefficients. 
The NNP-predicted diffusion coefficients are in good agreement with available experimental data and showed improved accuracy compared with conventional classical MD results. 

\section{Methodology}

\subsection{NNP Generation}

NNP models are generated using the GeNNIP4MD package.~\cite{Matsumura_2025} 
The training dataset consists of atomic configurations together with their corresponding energies, forces, and virials calculated from DFT calculations. 
The NNPs are constructed within the Deep Potential (DP) framework using the DeepPot-SE descriptor~\cite{Zhang_2018} implemented in DeePMD-kit.~\cite{Wang_2018,Zeng_2023} 
The se\_e2\_a descriptor with a cutoff radius of 6.0 \AA\ is employed, and the embedding and fitting networks are set to be (25, 50, 100) and (240, 240, 240), respectively. 
The NNP model is trained for 500,000 steps and the learning rate decayed from $1.0 \times 10^{-3}$ to $1.0 \times 10^{-8}$ with 5,000 decay steps. 

\subsection{Dataset Creation}
\begin{figure*}
\includegraphics[width=15.0cm,clip]{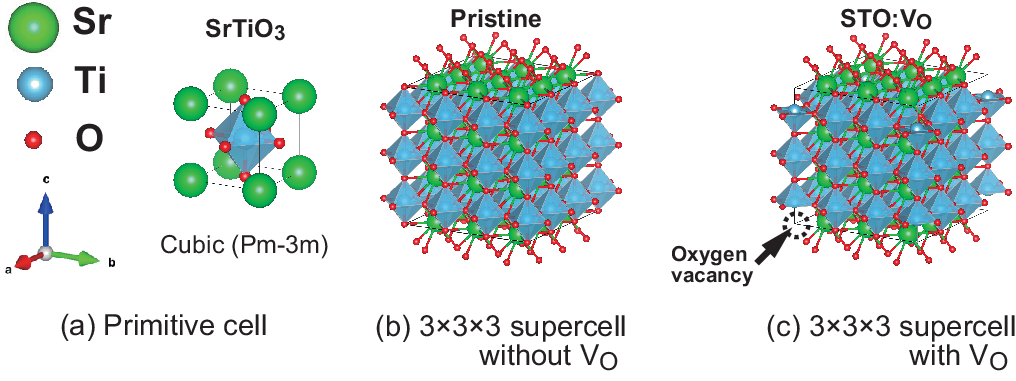}
\caption{
(a) Primitive cell of STO. 
(b) $3 \times 3 \times 3$ supercell model of pristine STO. 
(c) $3 \times 3 \times 3$ supercell model of STO:V$_{\rm O}$. 
The TiO$_6$ octahedral is highlighted in blue.
}
\label{fig:supercell}
\end{figure*}

To construct an initial dataset for NNP generation, DFT-based AIMD simulations are performed for both pristine STO and oxygen-deficient SrTiO$3$ containing a single oxygen vacancy (STO:V$_{\rm O}$). 
Figure~\ref{fig:supercell} schematically illustrates the $3 \times 3 \times 3$ supercell models employed for pristine STO and STO:V$_{\rm O}$. 
The supercell lattice constant is set to $A = 3a$, where $a$ denotes the experimental cubic lattice constant, $|\bm{a}_{1,2,3}| = 3.905$ \AA.~\cite{Madsen_1995,Ligny_1996,Cao_2000,Shimuta_2002} 
The pristine and oxygen-deficient supercells contain 135 and 134 atoms, respectively.

The AIMD simulations are carried out in the canonical NVT ensemble using a Nos\'e--Hoover thermostat~\cite{Nose_1984,Hoover_1985}, with the Nos\'e mass parameter set to 0.1. 
A time step of 1 fs is employed, and each simulation is run for 1 ps, corresponding to 1,000 MD steps. 
Supercell models with lattice constants of $1.00A$, $1.01A$, and $1.02A$ are prepared for both pristine STO and STO:V$_{\rm O}$. 
The temperature is initialized at 300 K and increased linearly to 2100 K at a rate of 1.8 K/fs to sample a wide range of thermally accessible atomic configurations. 
In addition, to sample further configurations at elevated temperatures, additional AIMD simulations are performed for STO:V$_{\rm O}$: simulations with a lattice constant of $1.01A$ are conducted at 1200 and 1400 K, while those with a lattice constant of $1.02A$ are conducted at 1600 and 1800 K. 
In total, 10,000 atomic configurations are generated from the pristine and oxygen-deficient systems, and the resulting dataset is divided into training and validation subsets in an 8:2 ratio.

All these AIMD calculations are performed using the Vienna {\it Ab initio} Simulation Package (VASP)~\cite{Kresse_1996a,Kresse_1996b} within the projector augmented-wave (PAW) method.~\cite{Blochl_1994,Kresse_1999} 
Exchange and correlation effects are described using the revised Perdew--Burke--Ernzerhof functional for solids (PBEsol).~\cite{PBEsol_2008} 
For PAW potentials, the core orbitals (number of valence electrons) are set to be Sr:[Ar]3$d^{10}$ (10), Ti:[Ne] (12), O:[He] (6). 
The plane-wave cutoff energy of 500 eV, the Gaussian smearing with the smearing width of 0.01 eV, the energy convergence criterion of 10$^{-6}$ eV per atom, and the Monkhorst-Pack $3 \times 3 \times 3$ $k$-point grids are used for the AIMD simulations of the $3 \times 3 \times 3$ supercell models. 

\subsection{Formation Energy of Oxygen Defect}

The defect formation energy of the oxygen vacancy is evaluated using the NNP constructed above. 
The formation energy is expressed as~\cite{Zhang_1991,Komsa_2012,Freysoldt_2014,Oba_2018}
\begin{equation} 
E_{\rm f} = \left\{ E_{\rm D} + E^{\rm corr} \right\} -E_{\rm P} + \mu_{\rm O} , 
\label{eq:Ef} 
\end{equation} 
where $E_{\rm D}$ is the total energy of STO:V$_{\rm O}$, $E_{\rm P}$ is the total energy of the corresponding pristine supercell, and $\mu_{\rm O}$ denotes oxygen chemical potential. 
For simplicity, the oxygen chemical potential is evaluated from the DFT total energy of an isolated oxygen molecule $E_{{\rm O}_2}$ as $\mu_{\rm O} = E_{{\rm O}_2}/2 = -5.14$ eV. 
The term $E^{\mathrm{corr}}$ accounts for finite-size effects arising from interactions between periodically repeated defects. 
For a neutral defect in a cubic supercell, the leading finite-size contribution is expected to scale as $L^{-3}$, where $L$ is the side length of the supercell.~\cite{Leslie_1985,Makov_1995,Fraser_1996,Dabo_2008} 

The formation energies are calculated using several supercell sizes for both pristine STO and STO:V$_{\rm O}$. 
For each supercell, the atomic positions are fully relaxed while keeping the lattice vectors fixed until the residual Hellmann–Feynman forces are less than 0.01 eV/\AA. 
The formation energy in the dilute limit is then estimated by fitting the calculated formation energies to $E_{\rm f}(L) = \alpha + \beta/L^3$ and taking the extrapolated value $\alpha$ as $L \rightarrow \infty$. 
To assess the accuracy of the NNP, the formation energies predicted by the NNP are calculated for the $2\times2\times2$, $3\times3\times3$, $4\times4\times4$, $5\times5\times5$, $6\times6\times6$, $8\times8\times8$, and $10\times10\times10$ supercells. 
For the $2\times2\times2$, $3\times3\times3$, and $4\times4\times4$ supercells, for which DFT calculations are performed, the NNP predictions are directly compared with the corresponding DFT results.

\subsection{Diffusion Coefficient of Oxygen Vacancies}

The diffusion of oxygen vacancies in STO is investigated by means of NNP-MD simulations. 
The mean-square displacement (MSD) of atomic species $A$ (= Sr, Ti, or O) is evaluated as 
\begin{equation} 
{\rm MSD}_{A}(t) = \frac{1}{N_A} \sum_i \left| \bm{r}_{Ai} (t) - \bm{r}_{Ai} (0) \right|^2 , 
\label{eq:MSD} 
\end{equation} 
where $N_A$ is the number of atoms of species $A$, and $\bm{r}_{Ai}(t)$ is the position of the $i$-th atom of species $A$ at time $t$. 
The diffusion coefficient of atomic species $A$, $D_A$, is obtained from the Einstein relation, 
\begin{equation} 
{\rm MSD}_{A}(t) = 6 D_{A} t + {\rm const.}  
\label{eq:DA} 
\end{equation} 
The diffusion coefficient of oxygen vacancies, $D_{\rm V}$, is then estimated from the oxygen diffusion coefficient $D_{\rm O}$ as 
\begin{equation} 
D_{\rm V} = \frac{1}{f} D_{\rm O} \frac{N_{\rm O}}{N_{\rm V}} , 
\label{eq:DV} 
\end{equation} 
where $f$ is the tracer correlation factor, taken to be 0.69 for vacancy-mediated oxygen diffusion on the anion sublattice of an $AB$O$_3$ perovskite, and $N_{\rm O}$ and $N_{\rm V}$ are the numbers of oxygen atoms and vacancies, respectively.~\cite{Souza_2012, Ishigaki_1988} 

To evaluate the oxygen-vacancy diffusion coefficient from the MSDs, we employ STO:V$_{\rm O}$ supercells substantially larger than those included in the training dataset. 
The simulation models are prepared as follows: a $10\times10\times10$ supercell of pristine STO containing 5,000 atoms is first constructed, and 15 oxygen atoms are then removed from randomly selected oxygen sites, corresponding to an oxygen-vacancy site fraction of $N_{\rm V}/(N_{\rm O}+N_{\rm V}) = 0.005$. 
The resulting STO:V$_{\rm O}$ supercell contains 4985 atoms. 
Five different STO:V$_{\rm O}$ configurations with different randomly selected oxygen-vacancy positions are prepared to improve the statistical reliability of the calculated diffusion coefficients. 

The NNP-MD simulations are performed using the Large-scale Atomic/Molecular Massively Parallel Simulator (LAMMPS) package~\cite{Thompson_2022} with a time step of 1 fs. 
Atomic trajectories are collected every 100 time steps, corresponding to an interval of 0.1 ps. 
For each STO:V$_{\rm O}$ configuration, an NPT simulation is first performed for 100 ps at the target temperature and a pressure of 1 bar. 
The temperature and pressure are controlled using a Nos\'e--Hoover thermostat~\cite{Nose_1984,Hoover_1985} and barostat, respectively. 
The equilibrium lattice constant is determined by averaging the supercell lattice constant over the final 50 ps of the NPT simulation. 

Subsequently, the supercell lattice constant is fixed at the corresponding average value, and an NVT simulation is performed at the target temperature. 
After an additional equilibration period of 20 ps in the NVT ensemble, atomic trajectories are collected for 200 ps and linearly fitted for the final 150 ps to evaluate the MSDs and diffusion coefficients. 
For each oxygen-vacancy configuration, three independent simulations with different initial velocity seeds are performed to further improve the statistical reliability. 
Consequently, at each temperature, the oxygen-vacancy diffusion coefficient is obtained as the average over a total of 15 independent simulations (5 oxygen-vacancy configurations $\times$ 3 initial velocity seeds). 

\section{Results and Discussion}

\subsection{Accuracy of NNP Model}

\begin{figure*}
\includegraphics[width=15.0cm,clip]{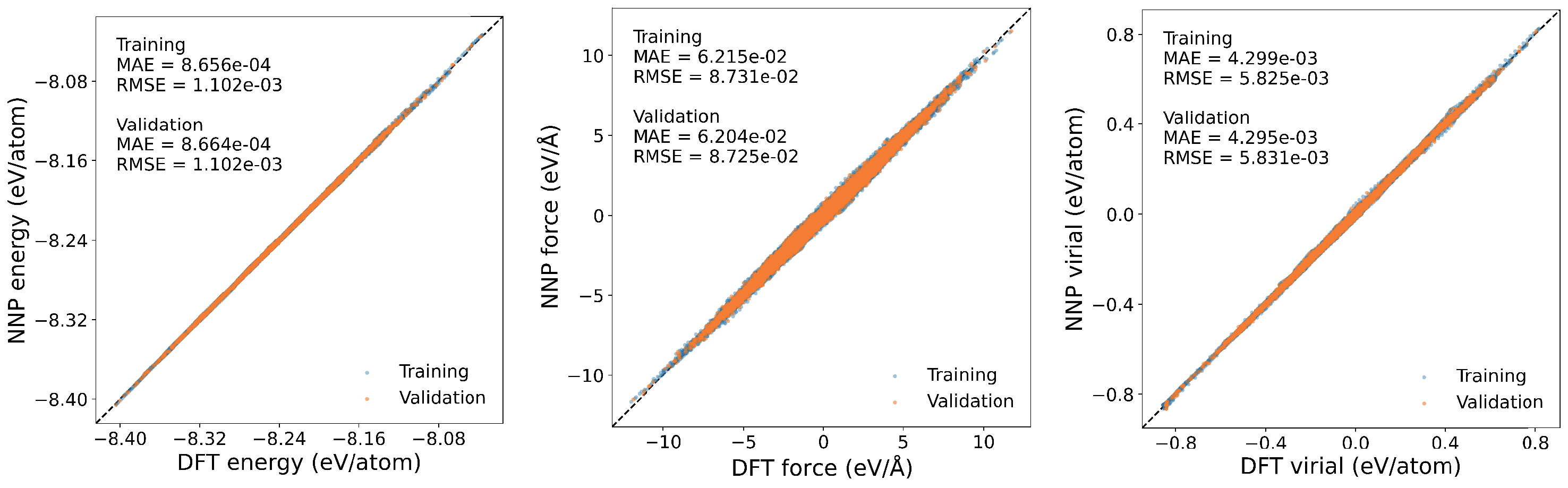}
\caption{
Parity plots of the NNP-predicted and DFT-calculated energies, forces, and virials. 
The solid diagonal line represents perfect agreement between the NNP and DFT results. 
}
\label{fig:pplot}
\end{figure*}

The accuracy of the NNP model is evaluated by comparing the predicted energies, atomic forces, and virials with the corresponding DFT reference values. 
Figure~\ref{fig:pplot} shows parity plots comparing the NNP-predicted and DFT-calculated energies, atomic forces, and virials. 
The root-mean-square error (RMSE) and mean absolute error (MAE) are evaluated for both the training and validation datasets. 
These results demonstrate that the trained NNP model reproduces the DFT reference energies, atomic forces, and virials with good accuracy. 
In particular, the accurate reproduction of atomic forces is important for NNP-MD simulations because the forces determine the local gradients of the potential-energy surface. 

\subsection{Oxygen-Vacancy Formation}

\begin{figure*}
\includegraphics[width=7.5cm,clip]{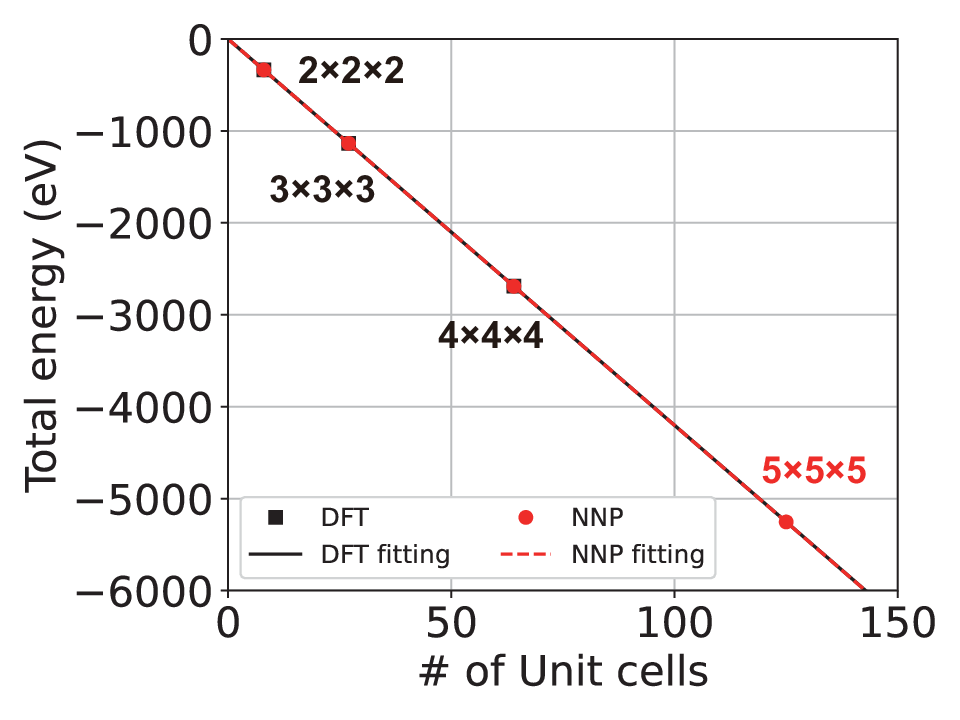}
\caption{
Total energies of pristine STO predicted by the NNP for the $2\times2\times2$, $3\times3\times3$, $4\times4\times4$, and $5\times5\times5$ supercells, together with the corresponding DFT results for the $2\times2\times2$, $3\times3\times3$, and $4\times4\times4$ supercells. 
The black solid and red dotted lines represent linear fits to the DFT and NNP total energies, respectively.
}
\label{fig:Etot}
\end{figure*}

Figure~\ref{fig:Etot} shows the NNP-predicted total energies of pristine STO for different supercell sizes, together with the corresponding DFT reference values. 
The NNP predictions are in good agreement with the DFT results for the $2 \times 2 \times 2$, $3 \times 3 \times 3$, and $4 \times 4 \times 4$ supercells. 
The slope of the linear fit, which corresponds to the total energy per unit cell, is $-42.03$ eV for the NNP results, in close agreement with the DFT value of $-42.02$ eV. 

\begin{figure*}
\includegraphics[width=7.5cm,clip]{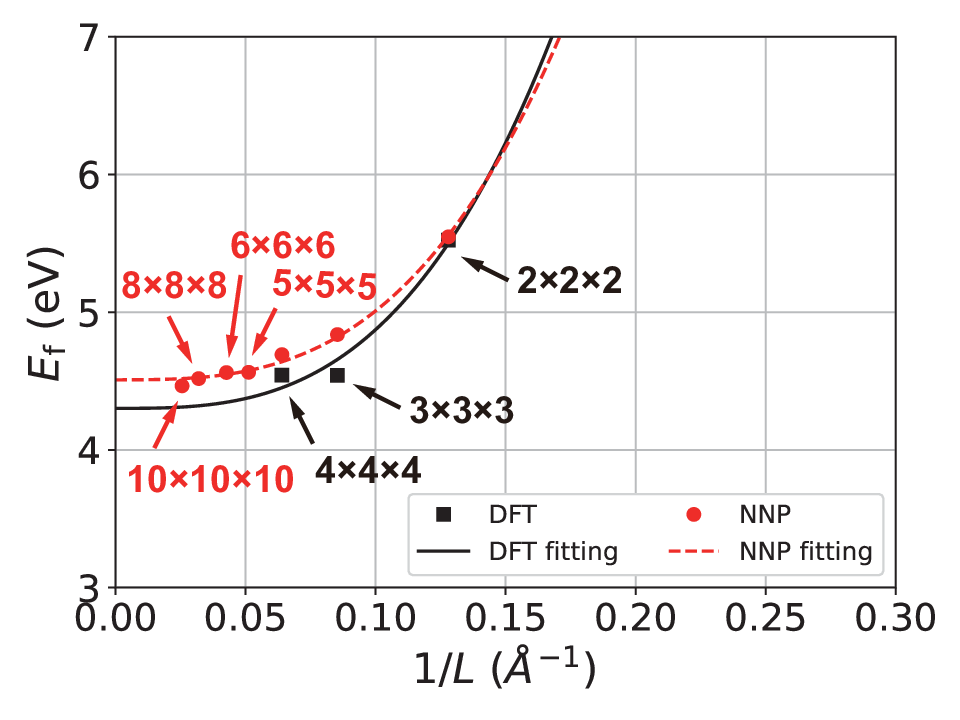}
\caption{
Oxygen-vacancy formation energies of STO:V$_{\rm O}$ predicted by the NNP for the $2\times2\times2$, $3\times3\times3$, $4\times4\times4$, $5\times5\times5$, $6\times6\times6$, $8\times8\times8$, and $10\times10\times10$ supercells, together with the corresponding DFT results for the $2\times2\times2$, $3\times3\times3$, and $4\times4\times4$ supercells. 
The black solid and red dotted lines represent the extrapolation fits to the DFT and NNP formation energies, respectively. 
}
\label{fig:Ef}
\end{figure*}

Figure~\ref{fig:Ef} shows the NNP-predicted oxygen-vacancy formation energies for different supercell sizes, together with the corresponding DFT reference values. 
The formation energies in the dilute limit are estimated by extrapolating the NNP and DFT results using the finite-size scaling relation for cubic (isotropic) supercells. 
The NNP predictions are also in good agreement with the corresponding DFT formation energies for the $2 \times 2 \times 2$, $3 \times 3 \times 3$, and $4 \times 4 \times 4$ supercells. 
The oxygen-vacancy formation energies in the dilute limit are estimated to be $4.30$ eV and $4.51$ eV from the NNP and DFT calculations, respectively. 
These results demonstrate that the present NNP model accurately reproduces the static energetics obtained from the DFT calculations. 

\subsection{Oxygen-Vacancy Diffusion}

\begin{figure*}
\includegraphics[width=7.5cm,clip]{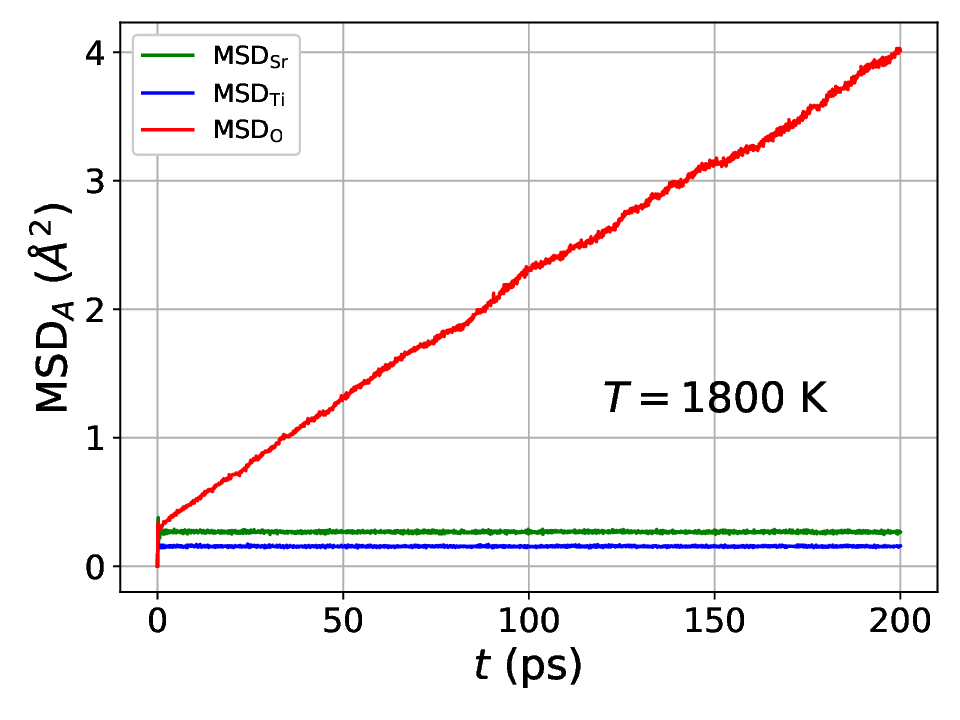}
\caption{
MSDs of Sr, Ti, and O atoms in a $10\times10\times10$ supercell of STO:V$_{\rm O}$ with an oxygen-vacancy site fraction of 0.005, obtained from the NNP-MD simulation at 1800 K.
}
\label{fig:msd}
\end{figure*}

The NNP-MD simulations of STO:V$_{\rm O}$ are performed to evaluate the MSDs of the constituent atomic species and the oxygen-vacancy diffusion coefficients. 
The constructed NNP model enables long-time simulations of large supercells that are computationally inaccessible to conventional AIMD simulations. 
Figure~\ref{fig:msd} shows the MSDs of Sr, Ti, and O atoms at 1800 K. 
The oxygen MSD increases approximately linearly in the long-time regime, indicating long-range oxygen migration mediated by oxygen vacancies. 
In contrast, the MSDs of Sr and Ti remain nearly constant, reflecting negligible long-range cation migration in the absence of cation vacancies.

\begin{figure*}
\includegraphics[width=7.5cm,clip]{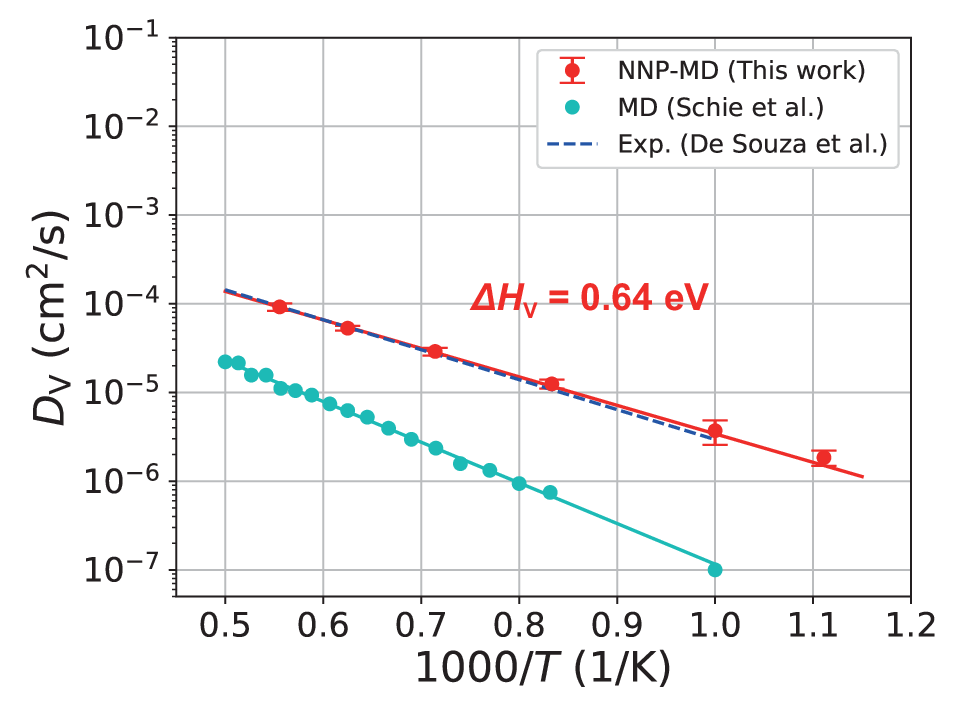}
\caption{
Temperature dependence of the oxygen-vacancy diffusion coefficient in STO obtained from the present NNP-MD simulations. 
The mean values obtained from 15 trajectories (five vacancy configurations $\times$ three initial velocity seeds) are shown at $T=1800$, 1600, 1400, 1200, 1000, and 900 K, with error bars representing the sample standard deviations. 
The oxygen-vacancy migration enthalpy is estimated to be $\Delta H_{\rm V}=0.64$ eV by fitting the calculated diffusion coefficients to the Arrhenius relation. 
Previous classical MD results~\cite{Schie_2012} and experimental data from De Souza {\it et al.}~\cite{Souza_2012} are included for comparison. 
}
\label{fig:DVO}
\end{figure*}

Figure~\ref{fig:DVO} shows the temperature dependence of the oxygen-vacancy diffusion coefficients obtained from the NNP-MD simulations. 
The previous classical MD results and experimental data are also included for comparison. 
The diffusion coefficients increase systematically with increasing temperature and follows Arrhenius behavior over the investigated temperature range. 
Fitting the calculated values to $D_{\rm V}=D_0\exp[-\Delta H_{\rm V}/(k_{\rm B}T)]$ gives a migration enthalpy $\Delta H_{\rm V}$.

The NNP-MD results show substantially better agreement with the experimental data than the classical MD results. 
In particular, the oxygen-vacancy migration enthalpy of $\Delta H_{\rm V}=0.64$ eV obtained from the present NNP-MD simulations is in good agreement with the experimental value of 0.67 eV,~\cite{Souza_2012} whereas the classical MD simulations yielded a larger value of 0.93 eV.~\cite{Schie_2012} 
The improved agreement indicates that the present NNP model provides a more accurate representation of the potential-energy landscape governing oxygen migration than the empirical potential used in the previous classical MD study. 
Because the tracer correlation factor is taken to be temperature independent, the numerical values may change the absolute scale of $D_{\rm V}$ but not the Arrhenius slope, namely, the migration enthalpy $\Delta H_{\rm V}=0.64$ eV.
The agreement in the migration enthalpy therefore provides a particularly robust validation of the temperature dependence predicted by the NNP-MD simulations. 

\section{Conclusion}

NNP-MD simulations of pristine STO and STO:V$_{\rm O}$ are performed to investigate oxygen-vacancy formation energetics and diffusion dynamics. 
The NNP model is trained using an AIMD dataset and accurately reproduces the corresponding DFT energies, atomic forces, and virials. 
The total energies of pristine STO and the oxygen-vacancy formation energies predicted for different supercell sizes are in good agreement with the DFT reference values. 
Extrapolation value in the dilute limit yields oxygen-vacancy formation energies of 4.30 and 4.51 eV from the NNP and DFT calculations, respectively. 
The oxygen-vacancy diffusion coefficients obtained from the NNP-MD simulations show substantially better agreement with the experimental data than those obtained from conventional classical MD simulations.
The calculated diffusion coefficients exhibit Arrhenius behavior with a migration enthalpy of 0.64 eV, in close agreement with the experimental value of 0.67 eV.

These results demonstrate that the present NNP model provides a reliable description of both the static energetics and diffusion dynamics of oxygen vacancies in STO. 
In particular, large-scale and long-time NNP-MD simulations enable the quantitative evaluation of oxygen-vacancy diffusion over length and time scales that are difficult to access using conventional AIMD simulations. 
The present approach therefore provides a useful framework for investigating point-defect formation and diffusion in STO and other functional materials.

%
%

\begin{acknowledgement}

K.N. gratefully acknowledges Yasufumi Sakai, Yuta Yoshimoto, Naoki Matsumura, Meguru Yamazaki, and Yuto Iwasaki of Fujitsu Limited for fruitful discussions. 
This work was partially supported by MEXT as ''Program for Promoting Researches on the Supercomputer Fugaku'' (Realization of innovative light energy conversion materials, Grant Number JPMXP1020210317), and JSPS KAKENHI Grant 22H00316. 
The numerical calculations were performed in part using the supercomputer facilities of the Institute for Solid State Physics, The University of Tokyo. 

\end{acknowledgement}

%
%
%


\bibliography{refs}

@article{Banerjee_2026,
	author = {Banerjee, Soumadip and Nishiguchi, Kazutaka and Ten-no, Seiichiro},
	doi = {10.1039/d6nj02761j},
	eprint = {https://pubs.rsc.org/nj/article-pdf/50/38/16464/14721649/d6nj02761j.pdf},
	issn = {1144-0546},
	journal = {New Journal of Chemistry},
	month = {10},
	number = {38},
	pages = {16464-16473},
	title = {Comparative theoretical study of layered perovskites Srn+1TinO3n+1 (n = 1, 2, 3) and SrTiO3 from photocatalytic perspectives},
	url = {https://doi.org/10.1039/d6nj02761j},
	volume = {50},
	year = {2026}}

@article{Ishigaki_1988,
	author = {Takamasa Ishigaki and Shigeru Yamauchi and Kohji Kishio and Junichiro Mizusaki and Kazuo Fueki},
	doi = {https://doi.org/10.1016/0022-4596(88)90067-9},
	issn = {0022-4596},
	journal = {Journal of Solid State Chemistry},
	number = {1},
	pages = {179-187},
	title = {Diffusion of oxide ion vacancies in perovskite-type oxides},
	url = {https://www.sciencedirect.com/science/article/pii/0022459688900679},
	volume = {73},
	year = {1988}}

@article{Nishiguchi_2025,
	author = {Nishiguchi, Kazutaka and Takayama, Tomoaki and Takasuka, Shogo and Harashima, Yosuke and Fujii, Mikiya and Ten-no, Seiichiro L.},
	doi = {10.1021/acs.jpclett.5c00070},
	issn = {1948-7185},
	journal = {The Journal of Physical Chemistry Letters},
	month = Mar,
	number = {11},
	pages = {2823--2830},
	publisher = {American Chemical Society (ACS)},
	title = {Divalent Cation Doping into SrTiO<sub>3</sub> for Enhancing the Photocatalytic Performance of Water Splitting},
	url = {http://dx.doi.org/10.1021/acs.jpclett.5c00070},
	volume = {16},
	year = {2025}}

@article{Souza_2012,
	author = {De Souza, Roger A. and Metlenko, Veronika and Park, Daesung and Weirich, Thomas E.},
	doi = {10.1103/PhysRevB.85.174109},
	issue = {17},
	journal = {Phys. Rev. B},
	month = {May},
	numpages = {11},
	pages = {174109},
	publisher = {American Physical Society},
	title = {Behavior of oxygen vacancies in single-crystal SrTiO${}_{3}$: Equilibrium distribution and diffusion kinetics},
	url = {https://link.aps.org/doi/10.1103/PhysRevB.85.174109},
	volume = {85},
	year = {2012}}

@article{Hoover_1985,
	author = {Hoover, William G.},
	doi = {10.1103/PhysRevA.31.1695},
	issue = {3},
	journal = {Phys. Rev. A},
	month = {Mar},
	numpages = {0},
	pages = {1695--1697},
	publisher = {American Physical Society},
	title = {Canonical dynamics: Equilibrium phase-space distributions},
	url = {https://link.aps.org/doi/10.1103/PhysRevA.31.1695},
	volume = {31},
	year = {1985}}

@article{Nose_1984,
	author = {Nos{\'e}, Sh{\=u}ichi},
	doi = {10.1080/00268978400101201},
	issn = {1362-3028},
	journal = {Molecular Physics},
	month = June,
	number = {2},
	pages = {255--268},
	publisher = {Informa UK Limited},
	title = {A molecular dynamics method for simulations in the canonical ensemble},
	url = {http://dx.doi.org/10.1080/00268978400101201},
	volume = {52},
	year = {1984}}

@article{Thompson_2022,
	author = {Aidan P. Thompson and H. Metin Aktulga and Richard Berger and Dan S. Bolintineanu and W. Michael Brown and Paul S. Crozier and Pieter J. {in 't Veld} and Axel Kohlmeyer and Stan G. Moore and Trung Dac Nguyen and Ray Shan and Mark J. Stevens and Julien Tranchida and Christian Trott and Steven J. Plimpton},
	doi = {https://doi.org/10.1016/j.cpc.2021.108171},
	issn = {0010-4655},
	journal = {Computer Physics Communications},
	pages = {108171},
	title = {LAMMPS - a flexible simulation tool for particle-based materials modeling at the atomic, meso, and continuum scales},
	url = {https://www.sciencedirect.com/science/article/pii/S0010465521002836},
	volume = {271},
	year = {2022}}

@article{Zeng_2023,
	author = {Zeng, Jinzhe and Zhang, Duo and Lu, Denghui and Mo, Pinghui and Li, Zeyu and Chen, Yixiao and Rynik, Mari{\'a}n and Huang, Li'ang and Li, Ziyao and Shi, Shaochen and Wang, Yingze and Ye, Haotian and Tuo, Ping and Yang, Jiabin and Ding, Ye and Li, Yifan and Tisi, Davide and Zeng, Qiyu and Bao, Han and Xia, Yu and Huang, Jiameng and Muraoka, Koki and Wang, Yibo and Chang, Junhan and Yuan, Fengbo and Bore, Sigbj{\o}rn L{\o}land and Cai, Chun and Lin, Yinnian and Wang, Bo and Xu, Jiayan and Zhu, Jia-Xin and Luo, Chenxing and Zhang, Yuzhi and Goodall, Rhys E. A. and Liang, Wenshuo and Singh, Anurag Kumar and Yao, Sikai and Zhang, Jingchao and Wentzcovitch, Renata and Han, Jiequn and Liu, Jie and Jia, Weile and York, Darrin M. and E, Weinan and Car, Roberto and Zhang, Linfeng and Wang, Han},
	doi = {10.1063/5.0155600},
	issn = {1089-7690},
	journal = {The Journal of Chemical Physics},
	month = aug,
	number = {5},
	publisher = {AIP Publishing},
	title = {DeePMD-kit v2: A software package for deep potential models},
	url = {http://dx.doi.org/10.1063/5.0155600},
	volume = {159},
	year = {2023}}

@article{Wang_2018,
	author = {Han Wang and Linfeng Zhang and Jiequn Han and Weinan E},
	doi = {https://doi.org/10.1016/j.cpc.2018.03.016},
	issn = {0010-4655},
	journal = {Computer Physics Communications},
	pages = {178-184},
	title = {DeePMD-kit: A deep learning package for many-body potential energy representation and molecular dynamics},
	url = {https://www.sciencedirect.com/science/article/pii/S0010465518300882},
	volume = {228},
	year = {2018}}

@article{Zhang_2018,
	author = {Linfeng Zhang and Jiequn Han and Han Wang and Saidi, {Wissam A.} and Roberto Car and Weinan E},
	issn = {1049-5258},
	journal = {Advances in Neural Information Processing Systems},
	language = {English (US)},
	note = {Publisher Copyright: {\textcopyright} 2018 Curran Associates Inc..All rights reserved.; 32nd Conference on Neural Information Processing Systems, NeurIPS 2018 ; Conference date: 02-12-2018 Through 08-12-2018},
	pages = {4436--4446},
	publisher = {Neural information processing systems foundation},
	title = {End-to-end symmetry preserving inter-atomic potential energy model for finite and extended systems},
	volume = {2018-December},
	year = {2018}}

@article{Kulichenko_2023,
	author = {Kulichenko, Maksim and Barros, Kipton and Lubbers, Nicholas and Li, Ying Wai and Messerly, Richard and Tretiak, Sergei and Smith, Justin S. and Nebgen, Benjamin},
	doi = {10.1038/s43588-023-00406-5},
	issn = {2662-8457},
	journal = {Nature Computational Science},
	month = mar,
	number = {3},
	pages = {230--239},
	publisher = {Springer Science and Business Media LLC},
	title = {Uncertainty-driven dynamics for active learning of interatomic potentials},
	url = {http://dx.doi.org/10.1038/s43588-023-00406-5},
	volume = {3},
	year = {2023}}

@article{van_der_Oord_2023,
	author = {van der Oord, Cas and Sachs, Matthias and Kov{\'a}cs, D{\'a}vid P{\'e}ter and Ortner, Christoph and Cs{\'a}nyi, G{\'a}bor},
	doi = {10.1038/s41524-023-01104-6},
	issn = {2057-3960},
	journal = {npj Computational Materials},
	month = sep,
	number = {1},
	publisher = {Springer Science and Business Media LLC},
	title = {Hyperactive learning for data-driven interatomic potentials},
	url = {http://dx.doi.org/10.1038/s41524-023-01104-6},
	volume = {9},
	year = {2023}}

@article{Podryabinkin_2023,
	author = {Podryabinkin, Evgeny and Garifullin, Kamil and Shapeev, Alexander and Novikov, Ivan},
	doi = {10.1063/5.0155887},
	issn = {1089-7690},
	journal = {The Journal of Chemical Physics},
	month = aug,
	number = {8},
	publisher = {AIP Publishing},
	title = {MLIP-3: Active learning on atomic environments with moment tensor potentials},
	url = {http://dx.doi.org/10.1063/5.0155887},
	volume = {159},
	year = {2023}}

@article{Novikov_2021,
	author = {Novikov, Ivan S and Gubaev, Konstantin and Podryabinkin, Evgeny V and Shapeev, Alexander V},
	doi = {10.1088/2632-2153/abc9fe},
	issn = {2632-2153},
	journal = {Machine Learning: Science and Technology},
	month = jan,
	number = {2},
	pages = {025002},
	publisher = {IOP Publishing},
	title = {The MLIP package: moment tensor potentials with MPI and active learning},
	url = {http://dx.doi.org/10.1088/2632-2153/abc9fe},
	volume = {2},
	year = {2021}}

@article{Zhang_2020,
	author = {Yuzhi Zhang and Haidi Wang and Weijie Chen and Jinzhe Zeng and Linfeng Zhang and Han Wang and Weinan E},
	doi = {https://doi.org/10.1016/j.cpc.2020.107206},
	issn = {0010-4655},
	journal = {Computer Physics Communications},
	pages = {107206},
	title = {DP-GEN: A concurrent learning platform for the generation of reliable deep learning based potential energy models},
	url = {https://www.sciencedirect.com/science/article/pii/S001046552030045X},
	volume = {253},
	year = {2020}}

@article{Sivaraman_2020,
	author = {Sivaraman, Ganesh and Krishnamoorthy, Anand Narayanan and Baur, Matthias and Holm, Christian and Stan, Marius and Cs{\'a}nyi, G{\'a}bor and Benmore, Chris and V{\'a}zquez-Mayagoitia, {\'A}lvaro},
	doi = {10.1038/s41524-020-00367-7},
	issn = {2057-3960},
	journal = {npj Computational Materials},
	month = jul,
	number = {1},
	publisher = {Springer Science and Business Media LLC},
	title = {Machine-learned interatomic potentials by active learning: amorphous and liquid hafnium dioxide},
	url = {http://dx.doi.org/10.1038/s41524-020-00367-7},
	volume = {6},
	year = {2020}}

@article{Vandermause_2020,
	author = {Vandermause, Jonathan and Torrisi, Steven B. and Batzner, Simon and Xie, Yu and Sun, Lixin and Kolpak, Alexie M. and Kozinsky, Boris},
	doi = {10.1038/s41524-020-0283-z},
	issn = {2057-3960},
	journal = {npj Computational Materials},
	month = mar,
	number = {1},
	publisher = {Springer Science and Business Media LLC},
	title = {On-the-fly active learning of interpretable Bayesian force fields for atomistic rare events},
	url = {http://dx.doi.org/10.1038/s41524-020-0283-z},
	volume = {6},
	year = {2020}}

@article{Smith_2018,
	author = {Smith, Justin S. and Nebgen, Ben and Lubbers, Nicholas and Isayev, Olexandr and Roitberg, Adrian E.},
	doi = {10.1063/1.5023802},
	issn = {1089-7690},
	journal = {The Journal of Chemical Physics},
	month = may,
	number = {24},
	publisher = {AIP Publishing},
	title = {Less is more: Sampling chemical space with active learning},
	url = {http://dx.doi.org/10.1063/1.5023802},
	volume = {148},
	year = {2018}}

@article{Piskunov_2004,
	author = {S Piskunov and E Heifets and R.I Eglitis and G Borstel},
	doi = {https://doi.org/10.1016/j.commatsci.2003.08.036},
	issn = {0927-0256},
	journal = {Computational Materials Science},
	number = {2},
	pages = {165-178},
	title = {Bulk properties and electronic structure of SrTiO3, BaTiO3, PbTiO3 perovskites: an ab initio HF/DFT study},
	url = {https://www.sciencedirect.com/science/article/pii/S0927025603001812},
	volume = {29},
	year = {2004}}

@article{Begum_2019,
	author = {Begum, Vijaya and Gruner, Markus Ernst and Pentcheva, Rossitza},
	doi = {10.1103/PhysRevMaterials.3.065004},
	issue = {6},
	journal = {Phys. Rev. Mater.},
	month = {Jun},
	numpages = {11},
	pages = {065004},
	publisher = {American Physical Society},
	title = {Role of the exchange-correlation functional on the structural, electronic, and optical properties of cubic and tetragonal ${\mathrm{SrTiO}}_{3}$ including many-body effects},
	url = {https://link.aps.org/doi/10.1103/PhysRevMaterials.3.065004},
	volume = {3},
	year = {2019}}

@article{Hou_2010,
	author = {Hou, Zhufeng and Terakura, Kiyoyuki},
	doi = {10.1143/jpsj.79.114704},
	issn = {1347-4073},
	journal = {Journal of the Physical Society of Japan},
	month = nov,
	number = {11},
	pages = {114704},
	publisher = {Physical Society of Japan},
	title = {Defect States Induced by Oxygen Vacancies in Cubic SrTiO3: First-Principles Calculations},
	url = {http://dx.doi.org/10.1143/JPSJ.79.114704},
	volume = {79},
	year = {2010}}

@article{Samanta_2012,
	author = {Samanta, Amit and E, Weinan and Zhang, S. B.},
	doi = {10.1103/PhysRevB.86.195107},
	issue = {19},
	journal = {Phys. Rev. B},
	month = {Nov},
	numpages = {14},
	pages = {195107},
	publisher = {American Physical Society},
	title = {Method for defect stability diagram from ab initio calculations: A case study of SrTiO${}_{3}$},
	url = {https://link.aps.org/doi/10.1103/PhysRevB.86.195107},
	volume = {86},
	year = {2012}}

@article{Katsumata_1998,
	author = {Tetsuhiro Katsumata and Yoshiyuki Inaguma and Mitsuru Itoh and Katsuyuki Kawamura},
	doi = {https://doi.org/10.1016/S0167-2738(98)00036-8},
	issn = {0167-2738},
	journal = {Solid State Ionics},
	number = {1},
	pages = {175-178},
	title = {Molecular dynamics simulation in SrTiO3},
	url = {https://www.sciencedirect.com/science/article/pii/S0167273898000368},
	volume = {108},
	year = {1998}}

@article{Schie_2012,
	author = {Schie, Marcel and Marchewka, Astrid and M{\"u}ller, Thomas and De Souza, Roger A and Waser, Rainer},
	doi = {10.1088/0953-8984/24/48/485002},
	journal = {Journal of Physics: Condensed Matter},
	month = {oct},
	number = {48},
	pages = {485002},
	publisher = {IOP Publishing},
	title = {Molecular dynamics simulations of oxygen vacancy diffusion in SrTiO3},
	url = {https://dx.doi.org/10.1088/0953-8984/24/48/485002},
	volume = {24},
	year = {2012}}

@article{Waldow_2016,
	author = {Waldow, Stephan P. and De Souza, Roger A.},
	doi = {10.1021/acsami.5b12574},
	eprint = {https://doi.org/10.1021/acsami.5b12574},
	journal = {ACS Applied Materials \& Interfaces},
	note = {PMID: 27121113},
	number = {19},
	pages = {12246-12256},
	title = {Computational Study of Oxygen Diffusion along a[100] Dislocations in the Perovskite Oxide SrTiO3},
	url = {https://doi.org/10.1021/acsami.5b12574},
	volume = {8},
	year = {2016}}

@article{Dawson_2020,
	author = {Dawson, James Alexander},
	doi = {https://doi.org/10.1002/pssb.201900422},
	eprint = {https://onlinelibrary.wiley.com/doi/pdf/10.1002/pssb.201900422},
	journal = {physica status solidi (b)},
	number = {1},
	pages = {1900422},
	title = {Dynamical Insights into Oxygen Diffusion in BaTiO3 and SrTiO3},
	url = {https://onlinelibrary.wiley.com/doi/abs/10.1002/pssb.201900422},
	volume = {257},
	year = {2020}}

@article{NNP_2007,
	author = {Behler, J\"org and Parrinello, Michele},
	doi = {10.1103/PhysRevLett.98.146401},
	issue = {14},
	journal = {Phys. Rev. Lett.},
	month = {Apr},
	numpages = {4},
	pages = {146401},
	publisher = {American Physical Society},
	title = {Generalized Neural-Network Representation of High-Dimensional Potential-Energy Surfaces},
	url = {https://link.aps.org/doi/10.1103/PhysRevLett.98.146401},
	volume = {98},
	year = {2007}}

@article{GAP_2010,
	author = {Bart\'ok, Albert P. and Payne, Mike C. and Kondor, Risi and Cs\'anyi, G\'abor},
	doi = {10.1103/PhysRevLett.104.136403},
	issue = {13},
	journal = {Phys. Rev. Lett.},
	month = {Apr},
	numpages = {4},
	pages = {136403},
	publisher = {American Physical Society},
	title = {Gaussian Approximation Potentials: The Accuracy of Quantum Mechanics, without the Electrons},
	url = {https://link.aps.org/doi/10.1103/PhysRevLett.104.136403},
	volume = {104},
	year = {2010}}

@article{SNAP_2015,
	author = {A.P. Thompson and L.P. Swiler and C.R. Trott and S.M. Foiles and G.J. Tucker},
	doi = {https://doi.org/10.1016/j.jcp.2014.12.018},
	issn = {0021-9991},
	journal = {Journal of Computational Physics},
	pages = {316-330},
	title = {Spectral neighbor analysis method for automated generation of quantum-accurate interatomic potentials},
	url = {https://www.sciencedirect.com/science/article/pii/S0021999114008353},
	volume = {285},
	year = {2015}}

@article{Andrade_2020,
	author = {Calegari Andrade, Marcos F. and Selloni, Annabella},
	doi = {10.1103/PhysRevMaterials.4.113803},
	issue = {11},
	journal = {Phys. Rev. Mater.},
	month = {Nov},
	numpages = {9},
	pages = {113803},
	publisher = {American Physical Society},
	title = {Structure of disordered ${\mathrm{TiO}}_{2}$ phases from ab initio based deep neural network simulations},
	url = {https://link.aps.org/doi/10.1103/PhysRevMaterials.4.113803},
	volume = {4},
	year = {2020}}

@article{Shimizu_2021,
	author = {Shimizu, Koji and Arguelles, Elvis F. and Li, Wenwen and Ando, Yasunobu and Minamitani, Emi and Watanabe, Satoshi},
	doi = {10.1103/PhysRevB.103.094112},
	issue = {9},
	journal = {Phys. Rev. B},
	month = {Mar},
	numpages = {10},
	pages = {094112},
	publisher = {American Physical Society},
	title = {Phase stability of Au-Li binary systems studied using neural network potential},
	url = {https://link.aps.org/doi/10.1103/PhysRevB.103.094112},
	volume = {103},
	year = {2021}}

@article{Chaney_2024,
	author = {Chaney, Gracie and Golov, Andrey and van Roekeghem, Ambroise and Carrasco, Javier and Mingo, Natalio},
	doi = {10.1021/acsami.4c02548},
	eprint = {https://doi.org/10.1021/acsami.4c02548},
	journal = {ACS Applied Materials \& Interfaces},
	note = {PMID: 38699998},
	number = {19},
	pages = {24624-24630},
	title = {Two-Step Growth Mechanism of the Solid Electrolyte Interphase in Argyrodyte/Li-Metal Contacts},
	url = {https://doi.org/10.1021/acsami.4c02548},
	volume = {16},
	year = {2024}}

@article{He_2022,
	author = {He, Ri and Wu, Hongyu and Zhang, Linfeng and Wang, Xiaoxu and Fu, Fangjia and Liu, Shi and Zhong, Zhicheng},
	doi = {10.1103/PhysRevB.105.064104},
	issue = {6},
	journal = {Phys. Rev. B},
	month = {Feb},
	numpages = {10},
	pages = {064104},
	publisher = {American Physical Society},
	title = {Structural phase transitions in $\mathrm{SrTi}{\mathrm{O}}_{3}$ from deep potential molecular dynamics},
	url = {https://link.aps.org/doi/10.1103/PhysRevB.105.064104},
	volume = {105},
	year = {2022}}

@article{Kresse_1996b,
	author = {Kresse, G. and Furthm\"uller, J.},
	doi = {10.1103/PhysRevB.54.11169},
	issue = {16},
	journal = {Phys. Rev. B},
	month = {Oct},
	numpages = {0},
	pages = {11169--11186},
	publisher = {American Physical Society},
	title = {Efficient iterative schemes for ab initio total-energy calculations using a plane-wave basis set},
	url = {https://link.aps.org/doi/10.1103/PhysRevB.54.11169},
	volume = {54},
	year = {1996}}

@article{Kresse_1996a,
	author = {G. Kresse and J. Furthm{\"u}ller},
	doi = {https://doi.org/10.1016/0927-0256(96)00008-0},
	issn = {0927-0256},
	journal = {Computational Materials Science},
	number = {1},
	pages = {15-50},
	title = {Efficiency of ab-initio total energy calculations for metals and semiconductors using a plane-wave basis set},
	url = {https://www.sciencedirect.com/science/article/pii/0927025696000080},
	volume = {6},
	year = {1996}}

@article{Kresse_1999,
	author = {Kresse, G. and Joubert, D.},
	doi = {10.1103/PhysRevB.59.1758},
	issue = {3},
	journal = {Phys. Rev. B},
	month = {Jan},
	numpages = {0},
	pages = {1758--1775},
	publisher = {American Physical Society},
	title = {From ultrasoft pseudopotentials to the projector augmented-wave method},
	url = {https://link.aps.org/doi/10.1103/PhysRevB.59.1758},
	volume = {59},
	year = {1999}}

@article{Blochl_1994,
	author = {Bl\"ochl, P. E.},
	doi = {10.1103/PhysRevB.50.17953},
	issue = {24},
	journal = {Phys. Rev. B},
	month = {Dec},
	numpages = {0},
	pages = {17953--17979},
	publisher = {American Physical Society},
	title = {Projector augmented-wave method},
	url = {https://link.aps.org/doi/10.1103/PhysRevB.50.17953},
	volume = {50},
	year = {1994}}

@article{PBEsol_2008,
	author = {Perdew, John P. and Ruzsinszky, Adrienn and Csonka, G\'abor I. and Vydrov, Oleg A. and Scuseria, Gustavo E. and Constantin, Lucian A. and Zhou, Xiaolan and Burke, Kieron},
	doi = {10.1103/PhysRevLett.100.136406},
	issue = {13},
	journal = {Phys. Rev. Lett.},
	month = {Apr},
	numpages = {4},
	pages = {136406},
	publisher = {American Physical Society},
	title = {Restoring the Density-Gradient Expansion for Exchange in Solids and Surfaces},
	url = {https://link.aps.org/doi/10.1103/PhysRevLett.100.136406},
	volume = {100},
	year = {2008}}

@article{Madsen_1995,
	author = {Lynnette D. Madsen and C.Niclas L. Johnson and Sissel N. Jacobsen and Ulf Helmersson and Staffan Rudner and Ivan Ivanov and Lars-David Wernlund and Lars Ryen and Eva Olsson},
	doi = {https://doi.org/10.1016/0167-9317(95)00129-8},
	issn = {0167-9317},
	journal = {Microelectronic Engineering},
	note = {1st European Meeting on Integrated Ferroelectrics},
	number = {1},
	pages = {123-127},
	title = {Structural and electrical characterization of sputter-deposited SrTiO3 thin films},
	url = {https://www.sciencedirect.com/science/article/pii/0167931795001298},
	volume = {29},
	year = {1995}}

@article{Ligny_1996,
	author = {de Ligny, Dominique and Richet, Pascal},
	doi = {10.1103/PhysRevB.53.3013},
	issue = {6},
	journal = {Phys. Rev. B},
	month = {Feb},
	numpages = {0},
	pages = {3013--3022},
	publisher = {American Physical Society},
	title = {High-temperature heat capacity and thermal expansion of ${\mathrm{SrTiO}}_{3}$ and ${\mathrm{SrZrO}}_{3}$ perovskites},
	url = {https://link.aps.org/doi/10.1103/PhysRevB.53.3013},
	volume = {53},
	year = {1996}}

@article{Cao_2000,
	author = {Cao, Lixin and Sozontov, E. and Zegenhagen, J.},
	doi = {https://doi.org/10.1002/1521-396X(200010)181:2<387::AID-PSSA387>3.0.CO;2-5},
	journal = {physica status solidi (a)},
	number = {2},
	pages = {387-404},
	title = {Cubic to Tetragonal Phase Transition of SrTiO3 under Epitaxial Stress: An X-Ray Backscattering Study},
	volume = {181},
	year = {2000}}

@article{Shimuta_2002,
	author = {Shimuta, Toru and Nakagawara, Osamu and Makino, Takahiro and Arai, Seiichi and Tabata, Hitoshi and Kawai, Tomoji},
	doi = {10.1063/1.1434547},
	eprint = {https://pubs.aip.org/aip/jap/article-pdf/91/4/2290/19278392/2290\_1\_online.pdf},
	issn = {0021-8979},
	journal = {Journal of Applied Physics},
	month = {02},
	number = {4},
	pages = {2290-2294},
	title = {{Enhancement of remanent polarization in epitaxial BaTiO3/SrTiO3 superlattices with ``asymmetric'' structure}},
	url = {https://doi.org/10.1063/1.1434547},
	volume = {91},
	year = {2002}}

@article{Matsumura_2025,
	author = {Matsumura, Naoki and Yoshimoto, Yuta and Yamazaki, Tamio and Amano, Tomohito and Noda, Tomoyuki and Ebata, Naoki and Kasano, Takatoshi and Sakai, Yasufumi},
	doi = {10.1021/acs.jctc.4c01613},
	eprint = {https://doi.org/10.1021/acs.jctc.4c01613},
	journal = {Journal of Chemical Theory and Computation},
	note = {PMID: 40195003},
	number = {8},
	pages = {3832-3846},
	title = {Generator of Neural Network Potential for Molecular Dynamics: Constructing Robust and Accurate Potentials with Active Learning for Nanosecond-Scale Simulations},
	url = {https://doi.org/10.1021/acs.jctc.4c01613},
	volume = {21},
	year = {2025}}

@article{Zhang_1991,
	author = {Zhang, S. B. and Northrup, John E.},
	doi = {10.1103/PhysRevLett.67.2339},
	issue = {17},
	journal = {Phys. Rev. Lett.},
	month = {Oct},
	numpages = {0},
	pages = {2339--2342},
	publisher = {American Physical Society},
	title = {Chemical potential dependence of defect formation energies in GaAs: Application to Ga self-diffusion},
	url = {https://link.aps.org/doi/10.1103/PhysRevLett.67.2339},
	volume = {67},
	year = {1991}}

@article{Komsa_2012,
	author = {Komsa, Hannu-Pekka and Rantala, Tapio T. and Pasquarello, Alfredo},
	doi = {10.1103/PhysRevB.86.045112},
	issue = {4},
	journal = {Phys. Rev. B},
	month = {Jul},
	numpages = {16},
	pages = {045112},
	publisher = {American Physical Society},
	title = {Finite-size supercell correction schemes for charged defect calculations},
	url = {https://link.aps.org/doi/10.1103/PhysRevB.86.045112},
	volume = {86},
	year = {2012}}

@article{Freysoldt_2014,
	author = {Freysoldt, Christoph and Grabowski, Blazej and Hickel, Tilmann and Neugebauer, J\"org and Kresse, Georg and Janotti, Anderson and Van de Walle, Chris G.},
	doi = {10.1103/RevModPhys.86.253},
	issue = {1},
	journal = {Rev. Mod. Phys.},
	month = {Mar},
	numpages = {53},
	pages = {253--305},
	publisher = {American Physical Society},
	title = {First-principles calculations for point defects in solids},
	url = {https://link.aps.org/doi/10.1103/RevModPhys.86.253},
	volume = {86},
	year = {2014}}

\end{document}